\documentclass[aps,preprint,nofootinbib]{revtex4}

\usepackage{graphicx}
\usepackage{epic}
\usepackage{eepic}
\usepackage{latexsym}

\newcommand{\eq}[1]{(\ref{#1})}
\newcommand{\be}{\begin{equation}}
\newcommand{\ee}{\end{equation}}
\newcommand{\bea}{\begin{eqnarray}}
\newcommand{\eea}{\end{eqnarray}}

\newcommand{\vs}[1]{\vspace{#1 mm}}
\newcommand{\hs}[1]{\hspace{#1 mm}}

\def\d{\delta}
\def\D{\Delta}

\def\fr{\frac}

\def\l{\lambda}

\def\m{\mu}
\def\n{\nu}

\def\r{\rho}

\def\th{\theta}

\def\del{\partial}

\let\bm=\bibitem
\def\nn{\nonumber}

\newcommand{\db}{\delta \hs{-.4}B}
\def\drb{\delta \hs{-.4}R_{1}}
\def\dri{\delta \hs{-.4}R_{2}}
\newcommand{\dth}{\delta \hs{-.4}\theta}

\begin{document}

\title{Brane Gases and Stabilization of Shape Moduli with\\
Momentum and Winding Stress}   

\author{Ali Kaya}
\email[]{ali.kaya@boun.edu.tr}
\affiliation{Bo\~{g}azi\c{c}i University, Department of
Physics, \\ 34342, Bebek, \.Istanbul, Turkey\vs{1.5}} 


\begin{abstract}
In a toy model with gases of membranes and strings wrapping over a
two-dimensional internal torus, we study the stabilization problem for
the shape modulus. It is observed that winding modes of partially
wrapped strings and momentum modes give rise to stress in the energy
momentum tensor. We show that this stress dynamically stabilizes the
shape modulus of the two torus.  
\end{abstract}

\maketitle

\section{introduction}

In usual Kaluza-Klein picture, the extra dimensions predicted by
string/M theory are assumed to be compact and small. 
The four dimensional low energy content of the theory (like the
spectrum of light particles, the strengths of the coupling constants
etc.) is determined by the internal space. Although phenomenologically
viable models can be obtained in Calabi-Yau or G2 manifold
compactifications, in its current formulation of string/M theory there
is no principle which would single out a specific compact
space. This is a major problem which prevents one to test the
validity of string/M theory by low energy phenomena.  

Besides the vacuum selection problem, one should also confront the
issue of the cosmological evolution of the extra dimensions. It is well
known that there is a limit to the change in the size of the internal
space following the epoch of  big bang nucleosynthesis
\cite{kolb}. Moreover, the size and the shape of the compact space are
parametrized by moduli scalar fields which are in general
dynamical. Therefore, to avoid any conflict with observations, moduli
fields should somehow be fixed in string/M theory.     

Brane gas cosmology, which was initiated in \cite{bv} and then
developed in \cite{bv2}, may offer a solution to the stabilization
problem in a cosmological context.\footnote{Different aspects of brane
  gas cosmology are studied in the literature: T-duality invariance, a
  possible loitering phase, generalization to curved Ricci flat
  manifolds and M-theory are discussed in the papers \cite{d1},
  \cite{d2}, \cite{d3} and \cite{d4}, respectively. Cosmological
  solutions with brane gas sources are constructed in
  \cite{s1,s2,s3,s4,s5,s6,s7}. Annihilation of winding branes which is
  crucial for the Brandenberger and Vafa decompactification mechanism
  is studied in \cite{r1,r2,r3,r4,r5,r6}. Inflation in the context of
  brane gas cosmology is considered in \cite{in1,in2} and the
stabilization of extra dimensions in a democratic wrapping scheme is
demonstrated in \cite{t1,t2,t3}.}   
In \cite{st1,st2,st3}, it is found that string winding and momentum
modes can stabilize the radii of an internal torus\footnote{See
  \cite{ms2} for some concerns about dilaton stabilization by string
  gases and \cite{ms3} for an alternative approach to
  stabilization.}. Moreover, as shown in \cite{stp1,stp2}, this
mechanism survives in the presence of linearized cosmological
perturbations.  In \cite{b1,b2} we generalize these results to higher
dimensional branes and find that the volume modulus of an internal
Ricci flat manifold can be stabilized by a gas of branes wrapping over
it. This extension is important for two reasons. Firstly winding
strings do not exist in the spectrum when the first homology class of
the internal space is zero and secondly they may not be
able stabilize volume modulus when the topology is not equal to the
product of 1-cycles. In \cite{b3}, we also show that long-wavelength
cosmological perturbations does not alter this result.     

In a recent paper \cite{ms1}, the stabilization of extra dimensions is
reconsidered in the presence of a string gas carrying two-form flux
and it is found that, in addition to volume modulus, the flux and the
shape moduli are also dynamically stabilized. In this paper, we address
the stabilization problem for the shape modulus without introducing
any flux. Following our earlier work \cite{b1}, we consider a toy
model in Einstein gravity with membrane and string gases wrapping over
an internal two-torus, but this time we take shape to be dynamical. In
that case, we observe that winding modes of partially wrapped branes
(strings in this model) and momentum modes give rise to stress in the
energy momentum tensor. We will show that this stress is capable of
dynamically stabilizing the shape modulus.  

As for the size moduli, we will see that when the observed space is
three dimensional the winding and the momentum modes stabilize them as
in \cite{b1}. When the dimension is greater than three, there appears
a complication which was pointed out earlier in \cite{s2},
i.e. partially wrapped  branes force the transverse compact directions
to expand. For stabilization of the size moduli in that case, one
should restrict the relevant physical parameters so that this does not
produce an instability. (In \cite{spyeni}, which appeared while our
work was being finalized, it is also found that special massless
string modes can stabilize volume, shape and dilaton moduli.)  

The organization of the paper is as follows. In the next section we
calculate the energy momentum tensors and obtain a background solution
corresponding to a fixed rectangular torus and an expanding observed
space. In section \ref{sec2}, we study the linearized fluctuations
around the background geometry and find that the shape modulus is
dynamically stabilized. In this section, we also discuss the
stabilization of the size moduli. In section \ref{sec3}, we conclude
with brief remarks and future directions. 
 
\section{Energy momentum tensors and the background solution \label{sec1}}

We consider an $(m+3)$-dimensional space-time with the metric  
\be
ds^2=-dt^2+e^{2B}\,dx^idx^i+G_{ab}\,dy^a dy^b,
\ee
where $i,j=1,..,m$ and $a,b=1,2$. The $m$-dimensional observed space
is spanned by $x^i$ and $y^a=(y^1,y^2)$ are coordinates on an internal
two-torus $T^2$ so that $y^a\sim y^a +2\pi$. In a cosmological setup
one should take  
\be
B=B(t),\hs{7}G_{ab}=G_{ab}(t).
\ee
A convenient parametrization of the metric on $T^2$ is 
\be
G_{ab}=\left[\begin{array}{cc}R_1^2& R_1R_2\,\sin\th\\R_1R_2\,\sin\th& R_2^2
\end{array}\right],
\ee
where $R_1$ and $R_2$ are the radii of $y^1$ and $y^2$, respectively,
and $\th$ is the shape modulus. Note that $\th=0$ corresponds to a
rectangular torus. We choose  
\be\label{on}
e^{\hat{a}}{}_{b}=\left[\begin{array}{cc} R_1\cos(\th/2)&R_2\sin(\th/2)\\
R_1\sin(\th/2)&R_2\cos(\th/2)\end{array}\right],
\ee
as an orthonormal basis in the tangent space (the hatted
and unhatted indices refer to the tangent and coordinate spaces,
respectively).   

Up to a certain constant prefactor, the energy of a membrane wrapping
over $T^2$ is equal to
\be\label{en1}
E_w=\sqrt{\det G}.
\ee
In the observed space the winding energy is concentrated at the position of
the membrane and this gives a delta function singularity. But for a gas of
such branes the delta function is smoothed out in the continuum fluid
approximation \cite{s2}. The energy spectrum of the momentum modes
corresponding to small vibrations of a membrane on $T^2$ is equal to 
the eigenvalues of the Laplacian of the metric $G_{ab}$ which are
characterized by two integers $n_a=(n_1,n_2)$. The energy of a mode
described by the wave-function $e^{iy^a n_a}$ is given by 
\be\label{en2}
E_m=\sqrt{G^{ab}n_a n_b}.
\ee
Note that $E_m$ is equal for two modes having quantum numbers $n_a$ and $-n_a$.

For strings wrapping over $y_1$ and $y_2$, the winding energy is equal to  
\be\label{es}
E_s=\sqrt{G_{11}}+\sqrt{G_{22}}.
\ee
As in the membrane case, the energy is distributed smoothly for a gas
in the continuum approximation. String momentum modes,\footnote{Here
  we consider semiclassical quantization of strings in a
  (non-covariant) physical gauge. Thus the tachyon is absent in the
  spectrum and there appears no critical dimension.} which are labeled
by $n_a=(n_1,n_2)$ with $n_1=0$ or $n_2=0$, can be viewed as a subset
of membrane momentum modes and their energy is still given by
\eq{en2}. The corresponding wave-functions are $e^{iy^1n_1}$ and
$e^{iy^2n_2}$, which represent small vibrations of strings wrapping
over $y^1$ and $y^2$, respectively. 

From \eq{en1}, \eq{en2} and \eq{es} the energy densities can be
 calculated as  
\be\label{ro}
\r=\frac{E}{\textrm{Vol}_S},
\ee
where ${\textrm{Vol}_S}=e^{mB}\sqrt{\det G}$ is the total spatial
volume. It is well known that (see e.g. \cite{rev}) one can write
a matter Lagrangian 
\be
{\cal L}_m=-2\r
\ee
for an energy density $\r$. Coupling ${\cal L}_m$ to Einstein-Hilbert
action 
\be
S=\int \sqrt{-g} \left[ R + {\cal L}_m \right],
\ee 
one obtains the field equations 
\be\label{fe}
R_{\m\n}-\fr{1}{2}g_{\m\n}R=T_{\m\n},
\ee
where 
\be
T_{\m\n}=-\fr{1}{\sqrt{-g}}\frac{\del}{\del g^{\m\n}}\left[\sqrt{-g}
{\cal L}_m\right] \label{emgenel}
\ee
is the energy momentum tensor. 

Using \eq{en1} in the above formulas, the energy momentum tensor for membrane 
winding modes can be found as  
\bea
T_{00}&=&N_w \,e^{-mB},\nn\\
T_{ij}&=&0,\label{emw}\\
T_{ab}&=&N_w\,e^{-mB}G_{ab}\nn,
\eea
where $N_w$ is a numerical factor containing the tension and the
number density of membranes. For string winding modes \eq{es} implies  
\bea
&&T_{00}=e^{-mB}\left[N_{s1}\fr{G_{11}}{\sqrt{\det G}}+N_{s2}
\fr{G_{22}}{\sqrt{\det G}}\right]\nn\\
&&T_{ij}=0,\label{sem}\\
&&T_{ab}=-e^{-mB}\left[N_{s1}\fr{G_{1a}G_{1b}}{\sqrt{G_{11}}\sqrt{\det G}}+
N_{s2}\fr{G_{2a}G_{2b}}{\sqrt{G_{22}}\sqrt{\det G}}\right]\nn,
\eea
where $N_{s1}$ and $N_{s2}$ are proportional to the tension and the
number densities of strings wrapping over $y^1$ and $y^2$,
respectively. Finally, the energy momentum tensor for a gas of string
and membrane momentum modes with fixed $n_a$ can be obtained from
\eq{en2} as  
\bea
T_{00}&=&N_m\,e^{-mB}\fr{\sqrt{G^{ab}n_an_b}}{\sqrt{\det G}},\nn\\
T_{ij}&=&0,\label{emm}\\
T_{ab}&=&N_m\,e^{-mB}\fr{n_a n_b}{\sqrt{\det G}\sqrt{G^{cd}n_cn_d}},\nn
\eea
where $N_m$ is the number density. 
Note that \eq{emm} obeys $g^{\m\n}T_{\m\n}=0$ and moreover $T_{ij}=0$, 
therefore momentum modes behave like a gas of massless particles
confined in the compact space. One can verify that all tensors
\eq{emw}, \eq{sem} and \eq{emm} are conserved  
\be
\nabla_{\m}T^{\m\n}=0,
\ee
so that the field equations \eq{fe} are consistent. 

It is clear from \eq{en1} and \eq{es} that winding modes become heavy as $T^2$
expands and in thermal equilibrium they are expected to decay into
lighter excitations. However, in usual brane gas scenario winding modes 
are assumed to fall out of thermal equilibrium, so one can take $N_w$ in
\eq{emw} and $N_{s1}$, $N_{s2}$ in \eq{sem} as constant numbers. 
Similarly, from \eq{en2} the momentum modes
become heavy as $T^2$ shrinks. Thus, for momentum modes to stop the
contraction of the internal space, there should be a mechanism for
them to survive the annihilation process. For a string gas 
(in dilaton gravity) the
annihilation rates of the winding and the momentum modes are equal to
each other by T-duality invariance of the field and the Boltzmann
equations \cite{r5}. Therefore T-duality dictates that
momentum modes should also fall out of thermal equilibrium like
winding modes. Although the toy model we consider does not have T-duality
invariance, the final scenario (obtained by including all 
possible excitations) is expected to have this symmetry
\cite{d1}. Viewing the model here as a part of the complete picture,
we assume that momentum modes also fall out of thermal equilibrium and
thus $N_m$ in \eq{emm} is a constant number density.    

It is possible to give an alternative argument without invoking 
T-duality invariance that momentum modes should exist in the
spectrum when the internal space contracts too much. Branes are
dynamical objects and one would expect them to fluctuate when they are
forced to collapse totally. This is crucial for stability since
otherwise any unwrapped brane propagating in the bulk would undergo a
complete collapse under the influence of its own tension. 

For momentum and (partially wrapped) string winding modes, \eq{sem}
 and \eq{emm} imply nonvanishing stress\footnote{This stress was not
 observed in \cite{b1} since only the volume modulus was taken to be
 dynamical there.} in the energy momentum tensor since in general
 $T_{\hat{a}\hat{b}}\not=0$ when $\hat{a}\not=\hat{b}$. It is
 important to refer to an orthonormal basis in determining the
 existence of stress. For instance, the off diagonal components of the
 winding energy momentum tensor of membranes \eq{emw} is non-vanishing
 in the coordinate basis. As we will see, however, (totally wrapped)
 membrane winding modes do not play a role in the 
 dynamics of the shape modulus. 

To obtain the background solution we assume
\be
B=B(t),\hs{10}G_{ab}=\textrm{const.}
\ee
Eq. \eq{fe} then yields 
\bea
&&m\ddot{B}+m\dot{B}^2=-\fr{(m-2)N_w}{(m+1)}\,\,e^{-mB}-\sum_i N_{m}^{(i)}\,
\fr{\sqrt{G^{ab}n_a^{(i)}n_b^{(i)}}}{\sqrt{\det G}}\,\,e^{-mB}\nn\\
&&\hs{22}-\fr{(m-1)N_{s1}}{(m+1)}\fr{\sqrt{G_{11}}}{\sqrt{\det G}}\,e^{-mB}-
\fr{(m-1)N_{s2}}{(m+1)}\fr{\sqrt{G_{22}}}{\sqrt{\det G}}\,e^{-mB},\label{feq1}\\
&&\ddot{B}+m\dot{B}^2=\fr{3N_w}{m+1}\,\,e^{-mB}
+\fr{2N_{s1}}{(m+1)}\fr{\sqrt{G_{11}}}{\sqrt{\det G}}\,e^{-mB}+
\fr{2N_{s2}}{(m+1)}\fr{\sqrt{G_{22}}}{\sqrt{\det G}}\,e^{-mB},\label{feq2}\\
&&\fr{(m-2)N_w}{(m+1)}\,G_{ab}+N_{s1}\fr{G_{1a}G_{1b}}{\sqrt{G_{11}}\sqrt{\det G}}+
N_{s2}\fr{G_{2a}G_{2b}}{\sqrt{G_{22}}\sqrt{\det G}}\nn\\
&&\hs{8}=\left[\fr{2N_{s1}}{(m+1)}\fr{\sqrt{G_{11}}}{\sqrt{\det G}}+
\fr{2N_{s2}}{(m+1)}\fr{\sqrt{G_{22}}}{\sqrt{\det G}}\right]G_{ab}+\sum_i \,N_{m}^{(i)}\,
\fr{n_a^{(i)}n_b^{(i)}}{\sqrt{G^{cd}n_c^{(i)}n_d^{(i)}}\sqrt{\det
    G}}\label{feq3},
\eea
where $N_{m}^{(i)}$ is the number density for the $i$'th momentum mode
with quantum numbers $n_a^{(i)}$. 

For a rectangular torus with $\th=0$ (i.e. $G_{12}=0$) the energies of
the modes with $(n_1,n_2)$ and $(n_1,-n_2)$ are equal to each
other. Therefore, these modes should have the same number
density. This implies that the off-diagonal component of \eq{feq3} is
identically satisfied  after the sum over the pairs $(n_1,n_2)$ and
$(n_1,-n_2)$. Taking moreover $R_1=R_2\equiv R_0$, the number
densities of strings should also be the same,
i.e. $N_{s1}=N_{s2}\equiv N_s$. Eq. \eq{feq3} then gives  
\be\label{r0}
\fr{2(m-2)N_w}{(m+1)}\,R_0^3+\fr{2(m-3)N_s}{(m+1)}\,R_0^2=\sum_i
\,N_{m}^{(i)}\,\sqrt{\d^{ab}n_a^{(i)}n_b^{(i)}}. 
\ee
Note that the cubic equation for $R_0$ has one unique positive root
for $m>2$, and for $m\leq 2$ the ansatz is not valid. From now on, we
restrict the dimension of the observed space $m>2$.  

Eq. \eq{feq2} can be solved to
get\footnote{By defining a new time coordinate it is possible to
  obtain the most general solution of \eq{feq2}, which asymptotically
  becomes \eq{bt}. Therefore, \eq{bt} describes the late time behavior
  we are interested in.} 
\be\label{bt}
B(t)=\fr{2}{m}\ln(bt),
\ee
where 
\be\label{bins}
b^2=\fr{3m\,N_w}{2(m+1)}+\fr{2m\,N_s}{(m+1)R_0}.
\ee
One can now check that \eq{feq1} is identically satisfied. Therefore
the metric  
\be\label{bs}
ds^2=-dt^2+(bt)^{4/m}dx^idx^i+R_0^2\,dy^ady^a
\ee
solves all field equations and describes a background supported by
brane winding and momentum modes. The moduli fields of this solution
are $R_1$, $R_2$ and $\th$, corresponding to the radii of $y^1$, $y^2$
and the angle between them. They have the vacuum values $R_1=R_2=R_0$ and
$\th=0$. From the energy-momentum tensors \eq{emw}-\eq{emm}, we see
that the winding and the momentum modes apply negative and positive
pressures along the compact directions, and they behave like
pressureless dust in the observed space. Therefore, it is not
surprising that in \eq{bs} the observed space expands exactly with the
same power for pressureless dust and the internal directions are
stabilized under the action of positive and negative pressures.  

\section{Stabilization of the moduli \label{sec2}}

Having obtained the background solution, we now study linearized
perturbations around it. The fluctuations are denoted by $\db$,
$\drb$, $\dri$ and $\dth$, which are taken to be the functions of
time. We carry the calculation in the orthonormal basis
\eq{on}. Variation of \eq{emw}, which is the winding energy momentum
tensor of membranes, gives  
\be\label{vw}
\d T_{\hat{0}\hat{0}}=-m\,N_w\,e^{-mB}\,\db,\hs{5}\d
T_{\hat{a}\hat{b}}=m\,N_w\,e^{-mB}\,\db\,\d_{ab}. 
\ee
From \eq{sem}, for string winding modes we find   
\bea
&&\d T_{\hat{0}\hat{0}}=-\fr{2m\,N_s}{R_0}\,e^{-mB}\,\db-\fr{N_s}{R_0^2}\,e^{-mB}\left[\drb+\dri\right],\nn\\
&&\d T_{\hat{a}\hat{b}}=\fr{m\,N_s}{R_0}\,e^{-mB}\,\db\,\d_{ab}+\fr{N_s}{R_0^2}\,e^{-mB}\,\left[\drb\,\d_{a2}\,\d_{b2}+\dri\,\d_{a1}\,\d_{b1}\right]\nn\\
&&\hs{9}-\fr{N_s}{R_0}\,e^{-mB}\,\left[\d_{a1}\,\d_{b2}+\d_{a2}\,\d_{b1}\right]\,\dth.
\eea
On the other hand, the perturbation of \eq{emm} for momentum modes yields  
\bea
&&\d
T_{\hat{0}\hat{0}}=-m\,\fr{|\vec{n}|N_m}{R_0^3}\,e^{-mB}\,\db-
\fr{N_m}{R_0^4\,|\vec{n}|}\,e^{-mB}\,n_a\,n^b\,\d 
e^{a}{}_{b}-\fr{|\vec{n}|N_m}{R_0^4}\,e^{-mB}\,\d
e^{a}{}_{a}\label{vm0}\\ 
&&\d T_{\hat{a}\hat{b}}=-\fr{N_m}{R_0^4\,|\vec{n}|}\,n_a\,n_b\,e^{-mB}\,
\left[m\,R_0\,\db+\d e^{c}{}_{c}\right]
+\fr{N_m}{R_0^4\,|\vec{n}|^3}\,n_a\,n_b\, e^{-mB}\,n_c\,n^d\, \d e^{c}{}_{d}\nn\\
&&\hs{13}
-\fr{2N_m}{R_0^4\,|\vec{n}|}\,n_c\,n_{(b}\,\d e^{c}{}_{a)},\label{vm1}
\eea
where 
\be
\d e^{a}{}_{b}=\left[\begin{array}{cc} \drb &R_0\dth/2\\
R_0\dth/2&\dri\end{array}\right],
\ee
$|\vec{n}|=\sqrt{n_1^2+n_2^2}$ and all operations on indices are
performed with the Kronecker delta function. Note that \eq{vm0} and
\eq{vm1} are invariant under $n_a\to-n_a$.  

The number densities of momentum modes are expected to be distributed
according to Boltzmann weight. This does not contradict with the
assumption that they fall out of thermal equilibrium, since just
before decoupling they are expected to obey Boltzmann distribution and
then evolve accordingly. Therefore it is sufficient to consider the
modes having minimum energy, since the contributions of others are
exponentially suppressed. In our case the minimum energy excitations
have quantum numbers $n_a=(1,0)$, $n_a=(-1,0)$, $n_a=(0,1)$ and
$n_a=(0,-1)$, which should have the same density $N_m$ since they have
the same energy. Note that both strings and membranes give rise to
such momentum modes and $N_m$ denotes the total number density. 

Neglecting all other momentum modes, the common radius of the internal torus obeys   
\be\label{r00}
\fr{(m-2)N_w}{(m+1)}\,R_0^3+\fr{(m-3)N_s}{(m+1)}\,R_0^2=2 N_m. 
\ee
Rewriting the field equations in the form 
\be\label{nfe}
R_{\hat{\m}\hat{\n}}=T_{\hat{\m}\hat{\n}}-\eta_{\hat{\m}\hat{\n}}\,
\fr{T^{\hat{\l}}{}_{\hat{\l}}}{(m+1)} ,
\ee
and using \eq{vw}-\eq{vm1}, we find the following second order
equations 
\bea
&&\ddot{\db}+\fr{4}{t}\,\dot{\db}+\fr{2}{m\,t}\,\dot{\d Y}=
-\fr{m(3\,N_w+4\,N_s)}{(m+1)R_0\,(b\,t)^2}\,\db
-\fr{2N_s}{(m+1)R_0\,(b\,t)^2}\,\d Y, \label{lf1}\\
&&\ddot{\d Y}+\fr{2}{t}\,\dot{\d Y}=\left[
\fr{(m-3)N_s}{(m+1)\,R_0}-\fr{6N_m}{R_0^3}\right]\,\fr{1}{(b\,t)^2}\,\d Y,\label{lf2}\\
&&\ddot{\d Z}+\fr{2}{t}\,\dot{\d Z}=
-\left[\fr{N_s}{R_0}+\fr{2N_m}{R_0^3}\right]\,\fr{1}{(b\,t)^2}\,\d Z,\label{lf3}\\
&&\ddot{\dth}+\fr{2}{t}\,\dot{\dth}=-\left[\fr{2N_s}{R_0}+\fr{4N_m}{R_0^3}\right]\,
\fr{1}{(b\,t)^2}\,\dth,\label{lf4}
\eea
and the first order initial constraint 
\be
(m-1)\dot{\db}+\dot{\d Y}+\fr{(m-1)}{t}\,\db+
\fr{(m-2)}{m\,t}\,\d Y=0,\label{incon}
\ee
where
\bea
&&\d Y=\fr{1}{R_0}\left[\drb+\dri\right],\\
&&\d Z=\fr{1}{R_0}\left[\drb-\dri\right].
\eea
As a consistency check of these equations we verify that the time
derivative of \eq{incon} is identically satisfied upon using the field
equations \eq{lf1} and \eq{lf2}. Note that $\d Y$ can be
solved from \eq{lf2} which in turn can be used in \eq{lf1} to 
fix $\db$. The perturbations $\d Y$ and $\d Z$ are related to volume
$R_1R_2$ and fraction $R_1/R_2$ moduli.  

Let us first consider the equation \eq{lf4} for $\dth$. Writing 
\be
\dth=t^k
\ee
one finds that $k$ obeys  
\be\label{ind}
k^2+k+c=0,
\ee
where $c=2N_s/(b^2R_0)+4N_m/(b^2R_0^3)$. The roots are
\be \label{k}
k=-\fr12\pm i\, \D,
\ee
where 
\be\label{D}
\D=\fr{\sqrt{4c-1}}{2}.
\ee
Using the value of $b$ in \eq{bins}, we see that $\D$ is a positive
reel number for $m>2$. This gives two solutions  
\be
\dth=c_1 \fr{\cos(\D\ln t)}{\sqrt{t}}+c_2 \fr{\sin(\D\ln t)}{\sqrt{t}},
\ee
which shows that $\dth\to 0$ and the shape modulus is stabilized. 

For $\d Y$ and $\d Z$, one can repeat the same analysis and get
equation \eq{ind} for the corresponding power $k$, where the constant
$c$ is the negative of the total number multiplying $\d Y/t^2$ or $\d
Z/t^2$ in the right hand side of the linearized equation. It is
necessary for stabilization that $\textrm{Re}\,k \geq 0$ and by
\eq{ind} this happens when  
\be
c\geq 0.
\ee
From \eq{lf3}, we see that $c=2N_s/(b^2R_0)+4N_m/(b^2R_0^3)>0$ for $\d
Z$ and thus the fraction modulus $R_1/R_2$ is stabilized. Eq. \eq{lf2}
shows that the volume modulus $R_1R_2$ is stabilized when $m=3$ since
the corresponding constant $c=6N_m/(b^2R_0^3)>0$. To stabilize
$R_1R_2$ for $m>3$ one should impose that
$N_m>R_0^2(m-3)N_s/(6m+6)$. This condition arises due to an effect
which was pointed out in \cite{s2}, i.e. (depending on the dimension
of the observed space) partially wrapped branes force the transverse
compact dimensions to expand. The condition on the number densities
avoids the presence of an instability. Up to this complication, which
occurs when $m>3$, all moduli fields in this toy model are stabilized
by brane winding and momentum modes. Ignoring the string gas,
i.e. setting $N_s=0$, we find that all fields are still stabilized by
membrane winding and momentum modes. Neglecting, on the other hand,
the membrane gas by setting $N_w=0$, $\d Y$ and thus the volume
modulus becomes unstable. Focusing on the shape modulus, we see that
either the momentum or the winding stress is capable of dynamically
stabilizing $\dth$.  

Therefore we see that the negative terms which appear in the right
hand sides of \eq{lf2}-\eq{lf4} are responsible for the stabilization
of the moduli fields. It seems that the membrane winding modes
characterized by $N_w$ does not play a role in the stabilization since
$N_w$ does not appear in the right hand sides. This is only true at
the linearized level for $\drb$ and $\dri$, and the membrane winding
modes help for the stabilization of the radii nonlinearly
\cite{b1}. However the right hand side of \eq{lf4}, which is
$(\hat{\m},\hat{\n})=(\hat{1},\hat{2})$ component of \eq{nfe}, is
equal to the off-diagonal component of the energy momentum tensor in
the orthonormal basis $T_{\hat{1}\hat{2}}$. From \eq{emw}, we see that
$T_{\hat{1}\hat{2}}=0$ for membrane winding modes exactly. Therefore,
winding modes of totally wrapped branes do not play a role in the
dynamics of the shape modulus. The situation is different for
partially wrapped branes. From \eq{lf2}-\eq{lf4} we see that winding
string modes, which are characterized by $N_s$, work for the
stabilization of $\dth$ and $\d Z$, but they try to destabilize $\d Y$
when $m>3$. In the linearized approximation, the contributions of
momentum modes $n_a=(1,0)$, $n_a=(-1,0)$, $n_a=(0,1)$ and $n_a=(0,-1)$
are negative in the right hand sides of \eq{lf2}-\eq{lf4}, thus they  
try to stabilize all moduli fields. 

Although the higher level momentum modes are exponentially suppressed,
one can determine their effect on the stabilization of the shape
modulus. From \eq{vm1}, we find that  
\be
\d T_{\hat{1}\hat{2}}=-N_m\,\fr{(n_1^4+n_2^4)}{R_0^3|\vec{n}|^3}\fr{\dth}{(bt)^2}<0.
\ee
for the sum of the modes $(n_1,n_2)$ and $(n_1,-n_2)$. This term
should be added to the right hand side of \eq{lf4} and since it is
negative we see that the pairs of the modes $(n_1,n_2)$ and
$(n_1,-n_2)$ force $\dth$ to be fixed. 

\section{\label{sec3}Conclusions}

In this work, we study the stabilization problem for the shape modulus
in a toy model in Einstein gravity with gases of membranes and strings
wrapping over an internal two-torus. Based on the earlier results in
the literature (for strings \cite{st1,st2,st3} and for higher
dimensional branes \cite{b1,b2}), it is not surprising to see that the
size moduli are stabilized. Our new observation is that the winding
modes of partially wrapped branes and the momentum modes give rise to
stress which is capable of dynamically stabilizing the shape
modulus. We also notice that winding modes of totally wrapped branes
do not play a role in the dynamics of the shape modulus. Of course the
model we consider is incomplete in many ways but it nicely illustrates
a generic dynamical behavior. Although for a two-torus there is only
one shape field to worry about, the mechanism can be generalized to
work for higher dimensional tori with more than one shape moduli .  

In this paper we carry the calculations in the framework of Einstein
gravity. It would be interesting to generalize these findings
to 10-dimensional dilaton gravity. In that case, the results would
depend on the brane type. In \cite{b2}, consistent field equations are
obtained for a D-brane gas coupled to dilaton gravity and a natural
extension of this work is to consider a dynamical shape in that toy
model.  

It is known that in string theory compactifications with fluxes, all
of the complex structure moduli including the shape of the internal
manifold can be stabilized. However, it turns out to be difficult to
stabilize the volume modulus in this framework. On the other hand,
volume stabilization can be achieved easily in brane gas
cosmology. This motivates the inclusion of brane gases carrying fluxes
and as shown in \cite{ms1} all moduli fields can be stabilized in a
simple toroidal compactification with a string gas and ambient
flux. In this work we show that the shape moduli can be stabilized
without introducing fluxes. However, one may need to introduce fluxes
for other moduli (like the ones related to form-fields generating
fluxes), therefore it would be interesting to develop further the
framework of brane gases with fluxes.   

We study the stabilization problem in a linearized setting. It would
be important to extend the results to the non-linear regime
without any apprxoximations. One
difficulty in this analysis is to determine the relative number densities
of the momentum modes. For instance, in a square two-torus with
$\th=0$, $n_a=(1,0)$ 
mode is lighter than $n_a=(1,1)$ mode. However, as $\th$ increases,
the energy of $n_a=(1,1)$ mode decreases. When $\sin\th=1/2$, the energies
of the modes $n_a=(1,0)$ and $n_a=(1,1)$ become equal. As $\th$
increases more, $n_a=(1,1)$ mode becomes the lighter one. (In
\cite{krd1} and \cite{krd2}, this relation between the shape of the
torus and the energy of the momentum modes was used to weaken the 
experimental bounds on theories with large extra dimensions). In a
non-linear analysis, fluctuations on $\th$ can be large and it is
difficult to determine the light species which would dominate the
dynamics.  

The torus is a special manifold where the eigenvalues of the Laplacian
are known explicitly. It is worth trying to extend the arguments where
this information is no longer available. In the realistic case, to any
topologically nontrivial cycle, one anticipates the existence of a
brane gas wrapping over that cycle. The results in brane gas cosmology
suggest that in such a scenario all moduli fields should be stabilized
by brane winding and momentum modes. 

\acknowledgments{This work is partially supported by Turkish Academy
  of Sciences via Young Investigator Award Program (T\"{U}BA-GEB\.{I}P).}

\end{document}